\documentclass[runningheads]{llncs}

\usepackage{amsmath}
\usepackage{amssymb}
\usepackage{booktabs}
\usepackage{graphicx}
\usepackage[colorlinks]{hyperref}
\usepackage[switch]{lineno}
\usepackage{subcaption}
\usepackage{longtable}
\usepackage[numbers,sort&compress]{natbib}

\usepackage{color}

\newcommand{\appendixref}[1]{\hyperref[#1]{Appendix~\ref*{#1}}}

\newcommand{\ignore}[1]{}
\newcommand{\OLD}{BST-19}
\newcommand{\OLDR}{BST-19$^*$}
\newcommand{\NEW}{STV-26} 

\usepackage[]{todonotes}

\newcommand{\lbelim}{l^\mathrm{elim}}
\newcommand{\lbquota}{l^\mathrm{quota}}
\newcommand{\lbdisp}{l^\mathrm{disp}}

\newcommand{\glb}{\textit{rul}}

\newcommand{\election}{\mathcal{E}}
\newcommand{\cands}{\mathcal{C}}
\newcommand{\winners}{\mathcal{W}}
\newcommand{\ballots}{\mathcal{B}}
\newcommand{\quota}{Q}
\newcommand{\seats}{N}
\newcommand{\order}{\pi}

\newcommand{\tally}{V}
\newcommand{\tv}{\tau}
\newcommand{\tallymax}[2]{\tally^{\mathrm{max}}_{#1,#2}}
\newcommand{\tallymin}[2]{\tally^{\mathrm{min}}_{#1,#2}}

\newcommand{\bvalue}{B}

\newcommand{\bvaluemin}[2]{\bvalue^{\mathrm{min}}_{#1,#2}}
\newcommand{\bvaluemax}[2]{\bvalue^{\mathrm{max}}_{#1,#2}}

\newcommand{\stand}{\mathcal{S}}

\newcommand{\tail}{\mathrm{tail}}
\newcommand{\plusplus}{\mathrel{||}}

\newcommand{\pile}[2]{\mathrm{pile}_{#1,#2}}

\title{Advances in STV Margin Computation\thanks{%
Accepted for \href{https://e-vote-id.org/}{E-Vote-ID 2026}.
This work was supported by the Australian Research Council
(Discovery Project DP220101012, OPTIMA ITTC IC200100009).}}

\author{
Michelle Blom  \inst{1} \orcidID{0000-0002-0459-9917} \and 
Alexander Ek\thanks{%
Funded by the European Union (ERC, CertiFOX, 101122653). Views and opinions expressed are however those of the author(s) only and do not necessarily reflect those of the European Union or the European Research Council. Neither the European Union nor the granting authority can be held responsible for them.
}       \inst{2}  \orcidID{0000-0002-8744-4805}  \and
Peter J. Stuckey   \inst{1}  \orcidID{0000-0003-2186-0459}  \and
Vanessa Teague \inst{3}  \orcidID{0000-0003-2648-2565} \and
Damjan Vukcevic    \inst{4}  \orcidID{0000-0001-7780-9586}
}

\authorrunning{Blom M, Ek A, Stuckey PJ, Teague V, Vukcevic D}

\institute{
Department of Data Science and AI, Monash University, Clayton, Australia \\
\email{michelle.blom@monash.edu}\and
Department of Computer Science, KU Leuven, Belgium\and
Thinking Cybersecurity Pty Ltd. \and
Department of Econometrics and Business Statistics, Monash University, Clayton,
Australia \\
}

\begin{document}

\maketitle

\begin{abstract}
\emph{Single transferable vote} (STV) is a multi-winner preferential
proportional electoral system. The \emph{margin} is the smallest number of
ballots that need to be manipulated to alter the set of winners. If we can
compute the margin of an STV election, or a reasonable lower bound on the
margin, we can use recent advances in auditing research to conduct a
risk-limiting audit of the election's winners. Knowledge of the margin also
provides insight into whether uncovered mistakes, or a known error rate in
ballot interpretation, could have influenced the outcome.  This paper
presents substantial improvements on an existing algorithm for computing
lower bounds on the margin of an STV election. These improvements allow us
to compute higher lower bounds for real STV elections, making
mismatch-based risk-limiting audits more practical.
\end{abstract}


\section{Introduction}

The \emph{single transferable vote} (STV), also called proportional ranked-choice voting, is an electoral system where voters rank candidates in order of preference and multiple candidates are elected in a manner reflecting voter preferences proportionally. STV is used for national, state, and local elections in Australia, for EU, national, and local elections in Ireland and Malta, and for some elections in Aotearoa/New Zealand, Northern Ireland, Scotland, and the USA.

STV tabulation proceeds in rounds, with each round either electing  or eliminating a candidate. In each case, every ballot sitting in the elected or eliminated candidate's tally pile is transferred to the next-highest-ranked eligible candidate on that ballot. When ballots are transferred from an elected candidate's tally, they are reduced in value according to a \emph{transfer value}.

A risk-limiting audit (RLA) is a process designed to efficiently provide affirmative
statistical evidence that the reported winners really won, and correct the
outcome (with a guaranteed high probability) if they did not win.

The \emph{margin}
is the minimum number of ballots that need to be altered---by
changing the marked preferences on the ballots---to change who wins.\footnote{%
A more general definition also allows alterations where ballots
can be  removed or added.  Here we only allow alterations that keep
the total number of ballots fixed.}
Determining the margin of an election informs us about how close it was, and allows us to conduct an RLA for the election using the mismatch-based method \citep{ek2025doing}.  Mismatch-based audits require only a lower bound, not an exact value, for the margin. However, if this bound is not tight then such audits may require sampling far more ballots than truly necessary. In this paper we present significant improvements in STV margin lower bounding, which should substantially reduce required sample sizes for mismatch-based STV audits.

 \citet{blom2019toward} presented a best-first branch-and-bound algorithm, which we will call \OLD{}, for
computing a lower bound on the margin of an STV election by searching over a tree of possible tabulation \textit{prefixes}. Each \textit{prefix} defines a sequence of seatings and eliminations over a number of tabulation rounds, starting with the first round $r = 1$ and ending at some round $r = n$ which may or may not be the last round of tabulation. This tree captured  outcomes that \textit{could} occur if the cast ballots were manipulated.
The algorithm used: a method for computing an upper bound on the margin; a linear relaxation of a mixed-integer non-linear program (MINLP) for computing a minimal manipulation to the ballots cast to realise an outcome that starts with a specific prefix; and heuristics for computing a lower bound on the number of ballots that would have to be altered to realise an outcome with a specific prefix. \OLD{} is currently the only approach developed for computing lower bounds on the margin of an STV election, while several approaches exist for computing upper bounds \citep{blom2020did,teague2022annexure}.

In this this paper, we improve \OLD{} in several ways:
\begin{itemize}
\item \emph{Solving of MINLPs} without resorting to linear relaxation.
\item \emph{More sophisticated lower-bounding heuristics} by considering future rounds and more sophisticated reasoning about transfer values.
\item \emph{Use of better margin upper bounds} from recent work \citep{blom2020did,teague2022annexure}.
\item \emph{Dominance checks} that allow us to discard prefixes from the search tree if something equivalent has been or is being explored.
\end{itemize}
Our improved algorithm for STV margins, which we will call \NEW{}, is able to find exact or near-exact margins for many real-world STV elections. For large 2-seat Australian STV elections, we are able to find substantially higher lower bounds on the margin than previously possible. For STV elections of the size commonly conducted in the USA, and at the local council level in Scotland, we can find margins that would allow us to conduct mismatch-based RLAs \citep{ek2025doing}.

An open source implementation is available in the following GitHub repository:
\url{https://github.com/michelleblom/pymarginstv}

\section{Background}
\subsection{Single Transferable Vote}
This paper considers the weighted inclusive Gregory method (WIGM) variant of STV. 
Voters rank candidates in order of preference, from first to last, in either a total order or leaving some candidates unranked, depending on the jurisdiction. Each ballot starts with a value of 1, which is subsequently reduced if used to elect a candidate to a seat. To be seated, a candidate's tally must reach or exceed the \textit{quota} $\quota$ (\textit{election threshold}).
The \textit{Droop quota} is typically used:
\begin{equation}
    \quota \ = \ \left\lfloor \frac{\text{\# of validly cast ballots}}{\text{\# of seats} + 1} \right\rfloor + 1.
    \label{eqn:Droop}
\end{equation}

Each candidate has a \emph{pile} of ballots, and  each ballot  a \emph{ballot value}. A candidate's \emph{tally} $\tally$ is the sum of the ballot values in their pile.
If a candidate's tally exceeds the quota, they are said to have a \emph{surplus}, and this is equal to the number of votes by which they exceed the quota, $\tally - \quota$.

Initially, each candidate is given all ballots on which they are ranked first. Tabulation proceeds in rounds in which a single candidate is seated or eliminated, until all seats are filled.
Ballots in a seated or eliminated candidate's pile are moved to other candidates' piles or discarded.
If 
the number of unfilled seats equals the number of remaining candidates, we seat all remaining candidates.

If no candidate has a quota at the start of a round, we eliminate the candidate with the lowest tally (breaking ties as defined by the jurisdiction), moving all ballots in their pile (at their current value) to the next most preferred eligible candidate on the ballot. If no such candidate exists, the ballot \textit{exhausts}. A \emph{remaining} candidate is one not yet seated or eliminated. An \textit{eligible}  candidate is a remaining candidate  that does not have a quota at the start of the round. 

When seating a candidate, the ballots in their pile continue in the tabulation, but are reduced in value according to their \emph{transfer value} $\tv$, which is the proportion of the tally that is due to the surplus, $\tv  = (\tally - \quota)/\tally$.
\ignore{
\begin{equation}
    \tv  = \ \frac{\tally - \quota}{\tally}.
    \label{eqn:TransferValue_simple}
\end{equation}
}
The new value of each ballot is equal to its current value multiplied by the transfer value. Each of these ballots is then given to the next most preferred eligible candidate on the ballot, or discarded if no such candidate exists. Ballots will skip any remaining candidate that had a quota at the start of the round, even if they have not yet been seated. 
Where multiple candidates achieve a quota simultaneously, they are seated in order of their tally, highest to smallest.

\begin{table}[t]
\caption{The 3-seat STV election from \autoref{eg:stv} between
candidates \texttt{A}--\texttt{E} and a quota of $\quota = \left\lfloor 1230/4 \right\rfloor +1 = 308$ votes.
(a)~Number of ballots of each type.
(b)~Tallies after each counting round, noting when quotas were
reached (in bold).}
\label{tab:EGSTV1}
    \begin{subtable}[t]{.30\columnwidth}
      \caption{}
      \label{tab:EGSTV1a}
      \centering
        \begin{tabular}{lr}
\toprule
Ranking & Count \\
\midrule
{}[\texttt{A}]       & 250 \\
{}[\texttt{B, A, C}] & 120 \\
{}[\texttt{C, D}]    & 400 \\
{}[\texttt{E}]       & 350 \\
{}[\texttt{C, E, D}] & 110 \\
\bottomrule
\end{tabular}
    \end{subtable}
    \begin{subtable}[t]{.70\columnwidth}
      \caption{}
      \label{tab:EGSTV1b}
      \centering
        \begin{tabular}{crrrr}
\toprule
Candidate & Round 1 & Round 2 & Round 3 & Round 4\\
\midrule
 & \texttt{C} elected    & \texttt{E} elected
 & \texttt{B} eliminated & \texttt{A} elected \\
 & $\tau_1 = 0.396$ & $\tau_2 = 0.12$  &  & \\
\midrule
\texttt{A} &         250  &         250    & 250    & \textbf{370}   \\
\texttt{B} &         120  &         120    & 120    &         ---    \\
\texttt{C} & \textbf{510} &         ---    & ---    &         ---    \\
\texttt{D} &           0  &         201.96 & 201.96 &         201.96 \\
\texttt{E} & \textbf{350} & \textbf{350}   & ---    &         ---    \\
\bottomrule
\end{tabular}
    \end{subtable}
\end{table}

\begin{example}
\label{eg:stv}
Consider the 3-seat STV election between candidates \texttt{A} to \texttt{E} in \autoref{tab:EGSTV1}, with 1230 validly cast ballots and a  quota of 308 votes.
The first preference tallies of \texttt{A} to \texttt{E} are 250, 120, 510, 0
    and 350 votes. Candidates \texttt{C} and \texttt{E} have a
    quota.
Candidate \texttt{C} has the largest surplus, at 202 votes, and is elected first.
Their transfer value is $\tau_1 = 202/510 = 0.396$.
The 400 [\texttt{C, D}] ballots are each given a weight of 0.396, and a total of 158.4 votes are added to \texttt{D}'s tally.
The 110 [\texttt{C, E, D}] ballots are each given a weight of 0.396, and are also given to candidate \texttt{D}, skipping \texttt{E} as they already have a quota.
Candidate \texttt{D} now has a tally of 201.96 votes.
Candidate \texttt{E} is then elected. Their transfer value would be $\tau_2 = 42/350 = 0.12$, but all of the ballots in their tally exhaust.
In the third round, no candidate has a quota's worth of votes, so the candidate
    with the smallest tally, \texttt{B}, is eliminated.
The 120 [\texttt{B, A, C}] ballots go to \texttt{A}, each retaining their current value of 1.
At the start of the fourth round, candidate \texttt{A} has reached a quota, at 370 votes, and is elected to the third and final seat. 
\end{example}

\subsection{Notation}

\begin{definition}[STV Election]
An \emph{STV election} is defined as a tuple $\election = (\cands, \ballots, \seats, \quota, \winners)$ where $\cands$ is the set of candidates up for election, $\ballots$ the multi-set of ballots cast in the election, $\seats$ the number of seats to be filled, $\quota$ the election quota (\autoref{eqn:Droop}), and $\winners$ the subset of candidates elected to a seat (the winners). 
Each ballot $b \in \ballots$ is a partial or complete ranking over the candidates $\cands$.
\end{definition}

\begin{definition}[Margin]
The \emph{margin}, or \emph{margin of victory}, for an STV election $\election = (\cands, \ballots,$ $ \seats, \quota, \winners)$ is defined as the smallest number of ballot manipulations required to ensure that candidates $\winners'  \neq \winners$ are elected to a seat (i.e., at least one candidate in $\winners'$ must not appear in $\winners$). A single manipulation changes the ranking on a single ballot $b \in \ballots$ to an alternate ranking.
\end{definition}

\begin{definition}[Election order]
Given an STV election $\election = (\cands, \ballots, \seats, \quota, \winners)$, we represent the outcome of $\election$ as an \emph{election order} $\order$, where $\order$ is a sequence of tuples ($c$, $a$) with $c \in \cands$, $a \in \{0,1\}$, and each $c \in \cands$ appears in at most one tuple. The tuple ($c$, 1) denotes that candidate $c$ is elected to a seat, while ($c$, 0) that $c$ has been eliminated. An order $\order$ is \emph{complete} if it involves the election of $\seats$ candidates, and \emph{partial} if fewer than $\seats$ candidates have been elected in $\order$.
\end{definition}

For example, the order $\order$ $=$ [(\texttt{A}, 0), (\texttt{C}, 1), (\texttt{B}, 0), (\texttt{D}, 1)] indicates that candidate \texttt{A} is eliminated in the first round of counting, \texttt{C} is next elected to a seat, \texttt{B} is then eliminated, and then \texttt{D} is elected to a seat.

An election $\election$ is tabulated in rounds.
We use $\stand_\order$ to denote the set of candidates that have not yet been elected or eliminated by the end of prefix $\order$ (i.e., the subset of candidates still standing at the end of $\order$), and $\stand_{\order,r}$ the subset of candidates still standing at the start of round $r$ of $\order$.

\subsection{Margin Lower Bound Computation with \OLD{}}\label{sec:OldMethod}
This section describes the overall structure of the STV margin lower bounding algorithm of \citet{blom2019toward}. We do not change this structure in this paper, but improve some of its components. The key components of the approach are: computation of an \emph{upper bound} on the margin; exploration of a tree of alternate election outcome \emph{prefixes}; heuristics for computing a lower bound on the manipulation required to realise outcomes that start with a specific prefix; and methods for pruning (ignoring) portions of this tree.

\begin{figure}[t]
\begin{tabbing}
\hspace{0.25in}\=\hspace{0.25in}\=\hspace{0.25in}\=\hspace{0.25in}\=\hspace{0.25in}\=\hspace{0.25in}\=\kill
\OLD($\election = (\cands, \ballots, \seats, \quota, \winners)$) \\
1 \> $\glb \leftarrow $ \textsc{ComputeUpperBound}($\election$) $\triangleright$ \autoref{sec:CSTVUB} \\
2 \> $F \leftarrow $ \textsc{InitialiseFrontier}($\election$)\\
3 \> {\bf while $F \neq \emptyset$ do} \\
4 \>\> $n, F \leftarrow $ Remove first node from $F$ \\
5 \>\> $F$, $\glb$ $\leftarrow$ \textsc{Expand}($n$, $F$, $\glb$, $\election$) \\ 
6 \> {\bf done} \\
7 \> {\bf return} $\glb$ \\
\\
\textsc{Expand}($n = (l, \order)$, $F$, $\glb$, $\election = (\cands, \ballots, \seats, \quota, \winners)$)\\
8 \> {\bf for each} remaining candidate $c$ {\bf do} \\
9 \>\> {\bf for each} action $a \in \{0, 1\}$ {\bf do} \\
10 \>\>\> $\order' \leftarrow \order \plusplus [(c, a)]$ $\triangleright$  Add $(c, a)$ to the end of $\order$\\
11 \>\>\> $l' \leftarrow$ Apply lower bounding heuristics to $\order'$ $\triangleright$ \autoref{sec:NewLBHeuristics} \\
12 \>\>\> {\bf if} $\order'$ is a complete outcome {\bf and} $l' < \glb$ {\bf then} \\
13 \>\>\>\>  $\glb \leftarrow l'$\\
14 \>\>\> {\bf else if} $\order'$ is partial outcome {\bf and} $l' < \glb$ {\bf then} \\
15 \>\>\>\> $F \leftarrow $ Insert $(l', \order')$ into $F$ \\
16 \> $F \leftarrow$ Remove all nodes $n'' = (l'', \order'')$ where $l'' \geq \glb$ \\
17\> {\bf return} $F, \glb$ 
\end{tabbing}
\vspace*{-5mm}
\caption{Key components of the \OLD{} STV margin lower bounding algorithm. Note that $\glb$ is a `running upper limit' on the margin \textit{lower bound} returned by the algorithm, as per \autoref{sec:OldMethod}, and not an upper bound on the margin itself.}
\label{alg:Structure}
\end{figure}

\autoref{alg:Structure} shows how these components fit together. \OLD{} used several heuristics to compute an upper bound on the STV margin (step 1).  We replace these with a heuristic that produces tighter upper bounds (see \autoref{sec:CSTVUB}). This upper bound is used to initialise a `running upper limit on the eventual lower bound', $\glb$, that the algorithm will return. \OLD{} maintains a frontier of nodes, $F$, each node $(l, \order) \in F$ composed of a lower bound $l$ and a prefix $\order$. The prefix is a partial or complete election order, and $l$ is a lower bound on the manipulation that would be required to realise an outcome that starts with $\order$. The frontier is initialised (step 2) by considering each candidate $c \in \cands$ and creating two nodes: one where they are seated and one where they are eliminated in the first round. When a node is created with prefix $\order$, we apply heuristics that compute a lower bound $l$ on the manipulation required to realise an outcome starting with $\order$ (see \autoref{sec:NewLBHeuristics}). If this lower bound is equal to or higher than our `running upper limit' ($\glb$), we ignore the node, otherwise we add it to our frontier. Nodes on the frontier are sorted in order of their  lower bound, smallest to largest.

While our frontier is non-empty, we remove the first node $n = (l, \order)$ and expand it (steps 4--5). When expanding a node, we create two child nodes for each remaining candidate $c$ (not yet seated or eliminated in $\order$). One child seats $c$ next and the other eliminates them (steps 9--10). If a child seats all of the reported winners of the election, we ignore it, as we are only interested in outcomes that \emph{change} who wins. We apply our lower bounding heuristics (see \autoref{sec:NewLBHeuristics}) to compute an $l$ for each child (steps 10--11). If the child is a complete outcome, filling all seats, and $l < \glb$, we update $\glb$ to $l$ (steps 12--13).
If the child is a partial outcome and $l < \glb$ we add it to our frontier (steps 14--15). Otherwise it is ignored. All nodes $n'' = (l'', \order'')$ on the frontier where $l'' \geq \glb$ are removed (step 16). Expansion updates the frontier and our `running upper limit' $\glb$.

When we have explored or ignored all nodes on the frontier, we return $\glb$ as the computed lower bound on the election $\election$'s margin (step 7). 

In the next section we describe how we have improved the computation of margin
upper bounds (\autoref{sec:CSTVUB}) and the lower bounding heuristics applied
to partial/complete outcomes (\autoref{sec:NewLBHeuristics}). We also add a
\emph{dominance check} that improves efficiency by completely ignoring some
nodes (\autoref{sec:NewPruningMethods}).

\section{Improvements}\label{sec:Improvements}

\subsection{ConcreteSTV Upper Bounds}\label{sec:CSTVUB}
\citet{blom2020did} and \citet{teague2022annexure} describe a method of
computing upper bounds on STV margins that searches for, and tests, actual manipulations of ballots, looking for those that result in changed
outcomes.
This heuristic, implemented in
\href{https://github.com/AndrewConway/ConcreteSTV}{ConcreteSTV}, finds better
(smaller) upper bounds than the heuristics used by \OLD.\footnote{%
See \appendixref{sec:concrete} for more details.}

\ignore{
We ran ConcreteSTV with settings that matched, as closely as possible, the variant chosen in this paper. However, it is always possible 
that slight differences mean that the bounds found with one version are not guaranteed to be valid in a different version.\footnote{For example,
the lower-bounding algorithm uses floating-point values; the ConcreteSTV implementation uses fixed-precision decimals with 6 decimal places. This is
unlikely to make any practical difference, but in principle it could.} 
The margin lower bounding algorithm that we present in this paper is not reliant on the correctness of provided upper bounds. If the provided upper bound is less than the true margin, the algorithm will return a lower bound that is less than or equal to it. This is still a valid lower bound on the margin, although the algorithm may have been able to find a better lower bound with a valid initial upper bound.
}

\subsection{New Lower Bound Heuristics}\label{sec:NewLBHeuristics}

In step 11 of \autoref{alg:Structure}, \OLD{} assigns to each prefix $\order$ the largest of several lower bounds: the
lower bound assigned to the parent of $\order$ (if $|\order| > 1$), an
\emph{elimination} lower bound; a \empty{quota} lower bound; or a lower bound
obtained by solving a relaxation of a MINLP. For a given $\order$, an optimal
solution to this MINLP was the smallest manipulation that could be made to the
cast ballots to realise an outcome that started with $\order$. The MINLP was
relaxed in two ways: by merging sequences of eliminated candidates into a
 `super candidate' that was eliminated in a single round; and linearising
all non-linearities. We use a similar MINLP in this paper, with the
super-candidate relaxation, but retain all non-linearities and use the WIGM
surplus transfer rules.\footnote{The original MINLP of \citet{blom2019toward}
used Australian Senate surplus transfer rules.} Our MINLP
now yields a \emph{lower bound} on the manipulation required to realise outcomes
starting with $\order$.\footnote{The formulation of this MINLP is provided in
\appendixref{app:minlp}.}

The remainder of this section is structured as follows. We first describe how relaxation is used to make the minimal manipulation MINLP tractable (\textit{Super candidate relaxation}). We then outline how we improve the existing lower bounding heuristics used by \OLD{}. We first explain our new approach for computing tighter lower and upper bounds on candidate tallies in varying contexts (\textit{Minimum and maximum tallies}), and then how these new bounds are used in \OLD{}'s \textit{elimination} and \textit{quota} lower bounds. Finally, we present our new lower bounding heuristic, the \textit{displacement lower bound}, which is used together with the elimination and quota lower bounds, and the minimal manipulation MINLP, to assign lower bounds to prefixes in step~11 of \autoref{alg:Structure}.

\subsubsection{Super candidate relaxation.}

Solving the minimal manipulation MINLP  becomes intractable when dealing with long election orders. We instead solve the MINLP for a relaxation of a given prefix $\order$, denoted $\tilde{\order}$, in which some of the sequences of eliminations present in $\order$ are grouped or merged. By reducing the total number of candidates in the election, by merging some candidates into a single `super candidate', the number of model variables is considerably reduced.

 Consider an election order $\order$ $=$ [($\texttt{A},$ 0), ($\texttt{C},$ 1), ($\texttt{B},$ 0), ($\texttt{E},$ 0), ($\texttt{F},$ 0) ($\texttt{D}$, 1)]. This order is relaxed  by grouping candidates $\texttt{B}$ and $\texttt{E}$ into one `super candidate' $\texttt{BE}$, producing $\tilde{\order}$ $=$ [($\texttt{A},$ 0), ($\texttt{C},$ 1), ($\texttt{BE},$ 0), ($\texttt{F},$ 0), ($\texttt{D}$, 1)]. Where ($\texttt{BE},$ 0) appears in the order, it represents candidates $\texttt{B}$ and $\texttt{E}$ being eliminated in some sequence.
  Formally, we apply candidate merging to  sequences of $n \geq 3$ candidate eliminations $c_1, \ldots, c_{n-1}, c_n$ by grouping candidates $c_1$ to $c_{n-1}$ into a `super candidate', leaving $c_n$ out of the merge. 
  When merging eliminated candidates, some constraints in the MINLP, concerned with ensuring those candidates have the lowest tally at the point of their elimination, are removed.
  We leave $c_n$ out of the merge because merging entire sequences of eliminated candidates produces a relaxation that is too aggressive, resulting in poor lower bounds on the margin.

\subsubsection{Minimum and Maximum Tallies.}

\OLD's lower bounding heuristics rely on computing minimum and maximum possible tallies for candidates in each round of a given prefix. If a ballot might have been distributed as part of a surplus transfer before arriving in a candidate $c$'s pile, overly conservative assumptions were used: it was assumed to have a value of 1 when calculating $c$'s maximum tally, and 0 when computing their minimum tally.

As a prefix does not stipulate the precise round that a seated candidate received their quota, and votes skip candidates that already have a quota during surplus distributions, there is some ambiguity as to which pile a vote belongs to in any given round. For prefix $\order$ and round $r$, we define the \emph{tail} of ballot $b = [x_1, \dots, x_m]$ as the order of remaining candidates that $b$ \emph{can} (but not necessarily will) be transferred through as the tabulation continues starting with round $r$.
\begin{equation}
    \tail_{\order,b,r} = \left[x_i \mid 1 \leq i \leq m \textrm{~and~} x_i \not\in \{c_{\order,1}, \dots, c_{\order,{r-1}}\} \right] \label{eq:tail}
\end{equation}
where  $c_{\order,i}$ is the candidate being elected or eliminated in position $i$ of order $\order$. 

\begin{example}
Consider prefix $\order = [(\texttt{A}, 1), (\texttt{B}, 1), (\texttt{C}, 1)]$. For the ballot $ b = [\texttt{A}, \texttt{B}, \texttt{C}]$, and round $r = 2$,  $\tail_{\order,b,r} = [\texttt{B}, \texttt{C}]$. For the ballot $b' = [\texttt{C}, \texttt{A}, \texttt{B}]$, $\tail_{\order,b',r} = [\texttt{C}, \texttt{B}]$.
\end{example}

The pile that ballot $b$ belongs to at the start of round $r$ will be one of the candidates in $\tail_{\order,b,r}$ or the exhausted pile (\textbf{ex}).
The knowledge of what happens in round $r$ (i.e., a candidate is seated or eliminated) gives extra context as to what pile a ballot $b$ could be in.
In particular, the only time piles become ambiguous is when two or more seatings occur in a row in $\order$.

We define $\pile{\order}{b,r}$ as the set of possible piles a ballot $b$ could be in at the start of round $r$ of  prefix $\order$. In the following, $a_{\order,i}$ is 0 when a candidate is eliminated in round $i$ of $\order$ and 1 if a candidate is elected. 
\begin{align}
\pile{\order}{b,r} \ &= \ \begin{cases}
    \{ \mathbf{ex} \} &\mbox{if } \tail_{\order,b,r} = \emptyset \\
    \{x_1, \dots, x_m, \mathbf{ex}\} &\mbox{if~} a_{\order,r-1} = a_{\order,r} = 1 \mbox{, where~} \tail_{\order,b,r} = [x_1, \dots, x_m]\\
    \{ x_1 \} &\mbox{otherwise, where~} \tail_{\order,b,r} = [x_1, \dots, x_m]
\end{cases} \label{eq:pile}
\end{align}

We can now define which ballots \textit{must} be in a given candidate's pile in a given round, and which ballots \textit{may} be in their pile.
\stepcounter{equation}
\begin{align}
\ballots^{\mathrm{must}}_{\order,c,r} \ &= \ \{ b \mid b \in \ballots \textrm{~where~} \{c\} = \pile{\order}{c,r}\} \tag{\theequation a} \label{eq:cand_pile_must} \\
\ballots^{\mathrm{maybe}}_{\order,c,r} \ &= \ \{ b \mid b \in \ballots \textrm{~where~} c \in \pile{\order}{c,r}\} \tag{\theequation b} \label{eq:cand_pile_maybe}
\end{align}

\begin{table}[t]
\centering
\caption{Example of how tails and piles evolve in \autoref{eg:tail-pile} for different ballots $b$ and rounds $r$ of a prefix $\order$, where \textbf{ex} denotes that a ballot has exhausted.}
\label{tab:tail_and_pile_example}
\begin{tabular}{llllll}
\toprule
        & & \multicolumn{4}{c}{prefix $\order = $ [(\texttt{A},0), (\texttt{B},1), (\texttt{C},1), (\texttt{D},0)]} \\
\cmidrule{2-6}
ballot $b$           &    & $r=1$ & $r=2$ & $r=3$ & $r=4$ \\
\midrule
\texttt{[A, D]}    & tail:   & \texttt{[A, D]} & \texttt{[D]} & \texttt{[D]} & \texttt{[D]} \\ 
                   & pile:   & \{\texttt{A}\} & \{\texttt{D}\} & \{\texttt{D}\} & \{\texttt{D}\} \\
\midrule
\texttt{[A, C, B]}  & tail:  & \texttt{[A, C, B]} & \texttt{[C, B]}  & \texttt{[C]}  & $\emptyset$ \\
                    & pile:  & \{\texttt{A}\} & \{\texttt{C}\} & \{\texttt{C}\} & \{\textbf{ex}\} \\
\midrule
\texttt{[A, B, C, D]} & tail: & \texttt{[A, B, C, D]} & \texttt{[B, C, D]} & \texttt{[C, D]} & \texttt{[D]} \\
                   & pile:   & \{\texttt{A}\} & \{\texttt{B}\} & \{\texttt{C, D, }\textbf{ex}\} & \{\texttt{D}\} \\
\bottomrule
\end{tabular}
\end{table}

\begin{example}
\label{eg:tail-pile}
 Consider the prefix $\order =$  [(\texttt{A},0),(\texttt{B},1),(\texttt{C},1),(\texttt{D},0)]. \autoref{tab:tail_and_pile_example} shows how $\tail_{\order,b,r}$ and $\pile{\order}{b,r}$ are computed for different ballots $b$ and rounds $r$ in $\order$. For ballot [\texttt{A}, \texttt{C}, \texttt{B}] and round 2, for example, the tail is [\texttt{C}, \texttt{B}] while the ballot can only be in one pile, that of candidate \texttt{C}.  The ballot [\texttt{A}, \texttt{C}, \texttt{B}] \emph{must} be in candidate \texttt{A}'s pile in round 1, and then in \texttt{C}'s pile in rounds 2 and 3. The ballot [\texttt{A}, \texttt{B}, \texttt{C}, \texttt{D}] \emph{must} be in candidate \texttt{A}'s pile in round 1, \texttt{B}'s pile in round 2, but then \emph{may} be in \texttt{C}'s, \texttt{D}'s, or the exhausted pile in round $3$.
Our determination of which pile a ballot could be in at round $r$ of $\order$ \textit{does not} consider events at rounds $r' > r$. 
\end{example}

We have not yet defined \emph{how much} a ballot $b$ contributes to the pile it is in.
As finding the pile of a ballot $b$ at the start of round $r$ of a prefix $\order$ is sometimes ambiguous, the value of a ballot $b$ is similarly sometimes ambiguous.
We denote $\bvaluemax{\order}{b,r}$ and $\bvaluemin{\order}{b,r}$ as the maximum and minimum possible value (between 0 and 1) of ballot $b$ at the start of round $r$ in prefix $\order$. Whenever a candidate $c$ is seated in a round $r$ of a prefix $\order$, there is an associated transfer value $T_{\order,c,r}$. To compute the minimum and maximum value of a ballot $b$ after it has passed through one or more surplus transfers, we need to establish lower and upper bounds on the transfer value associated with each of those transfers. Let $T^{\min}_{\order,c,r}$ and $T^{\max}_{\order,c,r}$ denote a lower and upper bound, respectively, on the transfer value for the seated candidate $c$ in round $r$ of $\order$ (defined in  \autoref{eq:transfervalue_minmax}).
\stepcounter{equation}
\begin{align}
\bvaluemax{\order}{b,1} \ &= \ \bvaluemin{\order}{b,1} \ = \ 1 \tag{\theequation a} \label{eq:bvalue_base} \\
\bvaluemax{\order}{b,r} \ &= \ \begin{cases}
\bvaluemax{\order}{b,r-1} \times T^{\max}_{\order,c,r-1} &\mbox{if~} b \in \ballots^{\mathrm{must}}_{\order,c,r-1} \mbox{~and~} a_{\order,r-1} = 1  \\
\bvaluemax{\order}{b,r-1} &\mbox{otherwise}
\end{cases} \tag{\theequation b} \label{eq:bvalue_max} \\
\bvaluemin{\order}{b,r} \ &= \ \begin{cases}
\bvaluemin{\order}{b,r-1} \times T^{\min}_{\order,c,r-1} &\mbox{if~} b \in \ballots^{\mathrm{maybe}}_{\order,c,r-1} \mbox{~and~} a_{\order,r-1} = 1 \\
\bvaluemin{\order}{b,r-1} &\mbox{otherwise}
\end{cases} \tag{\theequation c} \label{eq:bvalue_min}
\end{align}
When $\order$ contains no seatings, both the minimum and maximum value of a ballot in any round of $\order$ is $1$.
We compute minimum and maximum tallies as follows:
\begin{equation}
\tallymax{\order}{c,r} = \sum_{b \in \ballots^{\mathrm{maybe}}_{\order,c,r}} \bvaluemax{\order}{b,r},  \quad \quad
\tallymin{\order}{c,r} = \sum_{b \in \ballots^{\mathrm{must}}_{\order,c,r}} \bvaluemin{\order}{b,r} \label{eq:tally_minmax}
\end{equation}

We compute bounds on the transfer value for a candidate $c$, seated in round $r$ of a prefix $\order$, in \autoref{eq:transfervalue_minmax}.
\begin{equation}
T^{\max}_{\order,c,r} = \frac{\max\left(\quota,\, \tallymax{\order}{c,r}\right) - \quota}{\max\left(\quota,\, \tallymax{\order}{c,r}\right)}, \quad \quad
T^{\min}_{\order,c,r} = \frac{\max\left(\quota,\, \tallymin{\order}{c,r}\right) - \quota}{\max\left(\quota,\, \tallymin{\order}{c,r}\right)}  \label{eq:transfervalue_minmax}    
\end{equation}
As we are often computing bounds for prefixes that did not arise in practice, i.e., that do not follow from the cast ballots, candidates may be elected in positions without a quota. We take the max of quota and the actual tally in \autoref{eq:transfervalue_minmax}  to arrive at sensible transfer values in these contexts.

\begin{example}
Consider the example from \autoref{tab:EGSTV1} and now with the prefix $\order = [(\texttt{C},1),(\texttt{E},1),(\texttt{A},0)]$. At the start of the first round, all ballots sit in the pile of their highest ranked candidate, and have a value of 1. As there is no ambiguity around the location and value of ballots, $\ballots^{\mathrm{maybe}}_{\order,c,r=1} = \ballots^{\mathrm{must}}_{\order,c,r=1}$ and $\tallymin{\order}{c,r=1} = \tallymax{\order}{c,r=1}$ for all candidates $c$. Consequently, we can compute an exact transfer value for \texttt{C}, i.e., $T^{\min}_{\order,\texttt{C},r=1} = T^{\max}_{\order,\texttt{C},r=1} = 0.396$. At the start of the second round, candidate \texttt{E} will have a minimum tally of $\tallymin{\order}{\texttt{E},r=2} = 350$  votes and a maximum tally of $\tallymax{\order}{\texttt{E},r=2} = 393.56$. The difference arises as the 110 [\texttt{C}, \texttt{E}, \texttt{D}] votes sitting in \texttt{C}'s pile in round 1 may or may not skip over \texttt{E}, when transferred at a value of 0.396 each, depending on when \texttt{E} achieves their quota. We can compute lower and upper bounds on \texttt{E}'s transfer value in round 2 as follows.
\[
T^{\min}_{\order,\texttt{E},2} = \frac{\max(308, 350) - 308}
                                      {\max(308, 350)} = 0.12,
~~
T^{\max}_{\order,\texttt{E},2} = \frac{\max(308, 393.56) - 308}
                                      {\max(308, 393.56)}    = 0.22.
\]

Consider candidate \texttt{D}'s tally in $r=3$ of $\order$. Their minimum tally, $V^{\min}_{\order,\texttt{D},r=3}$, is equal to the sum of the minimum values of the ballots that \textit{must} be in their pile in round 3. These are the 400 $[\texttt{C},\texttt{D}]$ ballots, at  0.396 votes each,  and the 110 $[\texttt{C},\texttt{E},\texttt{D}]$ ballots, at $0.396 \times 0.12 = 0.04752$ votes each. Thus, $V^{\min}_{\order,\texttt{D},r=3} = 163.627$. The maximum tally of \texttt{D} in $r=3$ of $\order$ is equal to the sum of the maximum values of the ballots that \textit{maybe} in their pile in round 3 which, in this case, is the same set of ballots that \textit{must} be in their pile. By \autoref{eq:tally_minmax}, the maximum total value of the 400 $[\texttt{C},\texttt{D}]$ and 110 $[\texttt{C},\texttt{E},\texttt{D}]$ ballots in round 3 is $400 \times 0.396 = 158.4$ and $110 \times 0.396 = 43.56$. Note that the latter set of ballots is not in $\ballots^{\mathrm{must}}_{\order,\texttt{E},r=2}$. 
\[
V^{\min}_{\order,\texttt{D},r=3} = 163.627, \qquad
V^{\max}_{\order,\texttt{D},r=3} = 158.4 + 43.56 = 201.96
\]
The minimum and maximum tallies of \texttt{A} and \texttt{B} in round 3 of $\order$ are simpler to compute, as the ballots that \textit{must} and \textit{maybe} in their piles across rounds 1 to 3 are the same, and none of these ballots participate in a surplus transfer.
\[
V^{\min}_{\order,\texttt{A},r=3} = V^{\max}_{\order,\texttt{A},r=3} = 250, \qquad
V^{\min}_{\order,\texttt{B},r=3} = V^{\max}_{\order,\texttt{B},r=3} = 120 \label{eg:MinMax}
\]
\end{example}%

\subsubsection{Elimination lower bound.}
For prefix $\order$, the \emph{elimination lower bound}, $\lbelim_\order$, is a lower bound on the number of ballots we need to change to ensure that each eliminated candidate in $\order$ has the smallest tally in the round they are eliminated. 

For a candidate $c \in \cands$, eliminated in round $r$ of $\order$, we compute $c$'s minimum tally at that point, $V^{\min}_{\order,c,r}$, as per \autoref{eq:tally_minmax}. We also compute the maximum possible tally of each other \textit{remaining} candidate $c'$ (i.e., that is \textit{still standing}) at the start of round $r$, according to $\order$, as per \autoref{eq:tally_minmax}.  For $c$ to be eliminated in  $r$, we need their minimum tally at this point to be \textit{less} than the maximum tally of all  other candidates still standing. Otherwise, we need to take votes away from $c$ to make this so. If $c$'s minimum tally is \textit{greater} than the maximum tally of one of the candidates still standing, then they cannot possibly be eliminated in round $r$. Thus, we need to change \textit{at least} the following number of votes:
\begin{equation}
\lbelim_{\order,c} = \max_{c' \in \stand_{\order,r} \setminus \{c\} } \left(\frac{V^{\min}_{\order,c,r}  - V^{\max}_{\order,c',r}}{2}\right)^+
\label{eqn:LBElimOfE}
\end{equation}
where $(\cdot)^+$ is the positive part function.\footnote{%
This is defined as $(a)^+ = \max(a, 0)$, which has value $a$ if $a \geqslant 0$ and value 0 if $a < 0$.}
For each $c$ vs $c'$ comparison, the change involves giving some of the votes that would reside with $c$ to $c'$.  

This forms an elimination lower bound, $\lbelim_{\order,c}$, with respect to candidate $c$. The overall elimination lower bound for $\order$ is obtained by taking the maximum candidate-based elimination lower bound across all candidates eliminated in $\order$. Let $E_\order \subset \cands$ denote the set of candidates eliminated in order $\order$. This gives:
\begin{equation}
    \lbelim_{\order} = \max_{c \in E_\order} \lbelim_{\order,c}
    \label{eqn:LBElim}
\end{equation}

\begin{example}
Consider prefixes $\order = [(\texttt{C},1),(\texttt{E},1),(\texttt{A},0)]$ and $\order' = [(\texttt{C},1),(\texttt{B},1)]$  for the STV election of \autoref{tab:EGSTV1}. No candidate in the prefix $\order'$ has been eliminated, and so its elimination lower bound is 0.  In $\order$, \texttt{A} is eliminated in the third round. In this case, $\lbelim_{\order} = \lbelim_{\order,\texttt{A}}$ by \autoref{eqn:LBElim}, and
\begin{equation*}
\lbelim_{\order,\texttt{A}} = \max_{c' \in \{\texttt{B},\texttt{D}\}} \left(\frac{V^{\min}_{\order,\texttt{A},r=3}  - V^{\max}_{\order,c',r=3}}{2}\right)^+
\end{equation*}
by \autoref{eqn:LBElimOfE}. To compute $\lbelim_{\order}$, we need to compute: $V^{\min}_{\order,\texttt{A},3}$, $V^{\max}_{\order,\texttt{B},3}$ and $V^{\max}_{\order,\texttt{D},3}$. 

From \autoref{eg:MinMax}, $V^{\min}_{\order,\texttt{A},3} = 250$, $V^{\max}_{\order,\texttt{B},3} = 120$ and $V^{\max}_{\order,\texttt{D},3} = 201.96$. Consequently, $\lbelim_{\order,\texttt{A}}$ is the maximum of $0.5 \times (250 - 120) = 65$ and $0.5 \times (250 - 201.96) = 24.02$ which is 65 votes. Thus,  $\lbelim_{\order} = 65$.
\label{eg:ExampleELB}
\end{example}

\subsubsection{Quota lower bound.}

For a prefix $\order$, its \emph{quota lower bound} considers all the candidates that are seated in $\order$. Consider a candidate $c$ that is seated in round $r$ of $\order$. If the maximum tally of $c$ at that point is less than a quota, then $c$ cannot possibly have been seated and we need to give extra votes to $c$ to make it so. The quota lower bound with respect to candidate $c$ in $\order$, and the overall quota lower bound for $\order$ is given by:
\begin{equation}
    \lbquota_{\order,c} = \left( \quota - V^{\max}_{\order,c,r} \right)^+, \quad     \lbquota_{\order} = \max_{c \in W_\order} \lbquota_{\order,c}
    \label{eq:lbquota}
\end{equation}

\begin{example}
Consider the prefix $\order = [(\texttt{C},1),(\texttt{E},1),(\texttt{D},1)]$  for the STV election of \autoref{tab:EGSTV1}. Three candidates are elected: $\texttt{C}$ in the first round, $\texttt{E}$ in the second, and $\texttt{D}$ in the third. In this example, $W_\order = \{\texttt{C},\texttt{E},\texttt{D}\}$. In this case, 
\begin{equation*}
\lbquota_{\order,\texttt{C}}  =  \left( \quota - V^{\max}_{\order,\texttt{C},1} \right)^+, \quad \lbquota_{\order,\texttt{E}}  =  \left( \quota - V^{\max}_{\order,\texttt{E},2} \right)^+, \quad \lbquota_{\order,\texttt{D}} =  \left( \quota - V^{\max}_{\order,\texttt{D},3} \right)^+
\end{equation*}
$V^{\max}_{\order,\texttt{C},1}$ is equal to \texttt{C}'s first preference tally, 510, while $V^{\max}_{\order,\texttt{E},2}$ is 393.56 as per \autoref{eg:MinMax}. The prefix considered in \autoref{eg:MinMax} is that same as $\order$ here in rounds~1 and~2, but differs in round~3. To compute the maximum tally of \texttt{D} in round~3, we sum the maximum values of the ballots that \emph{may} be in their pile: the 400 $[\texttt{C},\texttt{D}]$ and 110 $[\texttt{C},\texttt{E},\texttt{D}]$. Note that the latter set of ballots are not in $\ballots^{\mathrm{must}}_{\order,\texttt{E},r=2}$. This gives us
$V^{\max}_{\order,\texttt{D},3} = 0.396 \times 400 + 0.396 \times 110 = 201.96$. Given $\quota = 308$, we have:
\begin{equation*}
\lbquota_{\order,\texttt{C}} = 0, \quad
\lbquota_{\order,\texttt{E}} = 0, \quad
\lbquota_{\order,\texttt{D}} = \left(308 - 201.96\right)^+ = 106.4
\end{equation*}
By \autoref{eq:lbquota}, $\lbquota_{\order} = 106.4$, the maximum of $\lbquota_{\order,\texttt{C}}$, $\lbquota_{\order,\texttt{E}}$, and $\lbquota_{\order,\texttt{D}}$.
\label{eg:ExampleQLB}
\end{example}

\subsubsection{Displacement lower bound.}

The elimination and quota lower bounds consider only the eliminations and seatings present in a prefix $\order$. If, by the end of  $\order$, no \textit{new} candidate has been elected, we know that \textit{something} must change in future rounds. Some reported loser must be elected in place of a reported winner. 

 Consider a prefix $\order$, concluding in round $r-1$, where it is clear that at least one reported loser still standing has to displace one of the reported winners still 
standing.
In this case, we need to ensure that at least
one of the reported losers will not be eliminated before one of the 
reported winners.

We compute the \emph{displacement lower bound} for $\order$, $\lbdisp_\order$, as follows. First, we check whether $\order$ already changes our reported outcome by  seating a reported loser  or eliminating a reported winner. In both cases, $\lbdisp_\order$ is zero. We then check whether there is scope to change who is elected in subsequent rounds, beyond $\order$. If the number of unfilled seats equals the number of subsequent rounds, all remaining candidates will be automatically seated, and $\lbdisp_\order$ is again zero. We then consider each reported loser $c$ that is still standing (not yet elected or eliminated) at the end of $\order$. We denote the set of such candidates as $\mathcal{L}_\order$. We compute for each $c \in \mathcal{L}_\order$, the cheapest way that we could elect $c$ in the future at the expense of a reported winner, $\lbdisp_{\order,c}$. The displacement lower bound we assign to $\order$, $\lbdisp_\order$, is the smallest of those computed for each reported loser $c \in \mathcal{L}_{\order}$. 
\begin{equation}
    \lbdisp_\order = \min_{c \in \mathcal{L}_\order} \lbdisp_{\order,c}
    \label{eq:LDISP}
\end{equation}
To compute the displacement lower bound for a candidate $c \in \mathcal{L}_\order$, we compute three values: the cheapest way we can make sure $c$ is not eliminated before some reported winner still standing ($\mathrm{DispCost}_{\order,c}$); the cheapest way we can ensure $c$ achieves a quota ($\mathrm{QuotaCost}_{\order,c}$); and the cheapest way we can ensure $c$ outlasts enough candidates to be automatically seated in the final round ($\mathrm{LeftAtEndCost}_{\order,c}$). The displacement lower bound for $c$ is then:
\begin{equation}
    \lbdisp_{\order,c} = \max\big\{ \mathrm{DispCost}_{\order,c},\, \min \{ \mathrm{QuotaCost}_{\order,c},\, \mathrm{LeftAtEndCost}_{\order,c}\} \big\}.
\end{equation}

To compute $\mathrm{DispCost}_{\order,c}$, we consider each reported winner $w$  still standing at the end of $\order$, $w \in \winners \cap \stand_\order$, and compute the minimum cost required to displace $w$ with $c$,  $\mathrm{DispCost}_{\order,c,w}$. $\mathrm{DispCost}_{\order,c}$ is the minimum of these displacement costs. 
\begin{equation}
    \mathrm{DispCost}_{\order,c} \gets \min_{w \in \winners \cap \stand_\order}\mathrm{DispCost}_{\order,c,w} 
\end{equation}
Recall that $r$ is the next round of tabulation after $\order$. We compute the maximum  tally $c$ could achieve from the end of $\order$ onward, in the context where $w$ is still standing, $\tallymax{\order}{c \prec w,r}$ (\autoref{eq:tallydp}), and  the minimum tally of candidate $w$ at the end of $\order$, $\tallymin{\order}{w,r}$ (\autoref{eq:tally_minmax}).  A lower bound on the cost of displacing $w$ with $c$ is equal to half the difference between these tallies.
\begin{equation}
        \mathrm{DispCost}_{\order,c,w} \gets \max \big\{0,\, \frac{1}{2}\big(\tallymin{\order}{w,r} - \tallymax{\order}{c \prec w,r} \big) \big\}
\end{equation}%
%
\begin{equation}
\tallymax{\order}{c \prec w,r} \ = \ \sum_{b \in \ballots} \begin{cases}
    \bvaluemax{\order}{b,r} &\mbox{if } c \prec w \mbox{ in } \tail_{\order,b,r} \\
    0 &\mbox{otherwise} 
\end{cases} \label{eq:tallydp}
\end{equation}
where $c \prec w$ in a list is true if $c$ appears before $w$ or if only $c$ appears.

For a candidate  $c$ to be seated, they must either achieve a quota or must never be eliminated. To achieve a quota, their maximum possible tally from $\order$ onward must reach a quota. Let $\tallymax{\order}{c\prec*,r}$ denote this maximum tally.
\begin{equation}
\tallymax{\order}{c \prec *,r} \ = \ \sum_{b \in \ballots} \begin{cases}
    \bvaluemax{\order}{b,r} &\mbox{if } c  \in \tail_{\order,b,r} \\
    0 &\mbox{otherwise} 
\end{cases} \label{eq:quotatallydp}
\end{equation}

We compute the cheapest way for $c$ to achieve a quota ($\mathrm{QuotaCost}_{\order,c}$), and then for $c$ to be automatically seated in the final round ($\mathrm{LeftAtEndCost}_{\order,c}$). 
\begin{equation}
    \mathrm{QuotaCost}_{\order,c} \gets \max \{0, \quota - \tallymax{\order}{c \prec *,r}\}
\end{equation}
For an election $\election$ with $N$ seats, and prefix $\order$,  candidate $c$ needs to not be eliminated before $L_{\election,\order} = |\stand_\order| - (N - |W_\order|)$ other  remaining candidates. We compute and sort the displacement costs between $c$ and each remaining alternate candidate, both reported losers and winners, and take the maximum of the first  $L_{\election,\order}$ of these displacement costs to form $\mathrm{LeftAtEndCost}_{\order,c}$. 

Let $\mathcal{D}^<_{\order,c}$ denote the set of minimum costs required to displace each remaining candidate $c' \in \stand_\order \setminus \{c\}$ with $c$, in ascending order, and $d_i$ the $i$th cost in $\mathcal{D}^<_{\order,c}$.
\begin{equation}
\mathcal{D}^<_{\order,c} \gets [\mathrm{DispCost}_{\order,c,c'} \mid c' \in \stand_\order \setminus \{c\}] \quad    \mathrm{LeftAtEndCost}_{\order,c} =  d_{L_{\election,c}}
\end{equation}

\begin{example}
Consider the prefix $\order = [(\texttt{C},1),(\texttt{E},1)]$ for the STV election of \autoref{tab:EGSTV1}. This $\order$ reflects the start of the reported outcome in which \texttt{C} and \texttt{E} are elected. In the reported outcome, \texttt{A} is elected in round 4. We know that at some point after $\order$ we will need to elect some candidate \textit{other} than \texttt{A}. The reported losers still standing at the end of $\order$ are \texttt{B} and \texttt{D}. To compute $\lbdisp_{\order}$ we need to compute a displacement lower bound for each of these reported losers, $\lbdisp_{\order,\texttt{B}}$ and $\lbdisp_{\order,\texttt{D}}$.
\[
    \lbdisp_{\order,\texttt{B}} = \max\big\{ \mathrm{DispCost}_{\order,\texttt{B}},\, \min \{ \mathrm{QuotaCost}_{\order,\texttt{B}},\, \mathrm{LeftAtEndCost}_{\order,\texttt{B}}\} \big\}.
\]
\[
    \mathrm{DispCost}_{\order,\texttt{B}} \gets \min_{w \in \{\texttt{A}\}} \max \big\{0,\, \frac{1}{2}\big(\tallymin{\order}{w,r=3} - \tallymax{\order}{\texttt{B} \prec w,r=3} \big) \big\}
\]
To compute $\mathrm{DispCost}_{\order,\texttt{B}}$, we need to compute $\tallymin{\order}{\texttt{A},r=3}$ and $\tallymax{\order}{\texttt{B} \prec \texttt{A},r=3}$. The only ballots that \texttt{A} must have in their pile at the end of $\order$ are  the 250 $[\texttt{A}]$ ballots. The only ballots that mention \texttt{B} are the 120 $[\texttt{B},\texttt{A},\texttt{C}]$ ballots.
\[
\tallymin{\order}{\texttt{A},r=3} = 250, \quad
\tallymax{\order}{\texttt{B} \prec \texttt{A},r=3} = 120, \quad 
 \mathrm{DispCost}_{\order,\texttt{B}} = \max \{0, \frac{1}{2} (250 - 120)\} = 65 
\]
A lower bound on the cost required to give \texttt{B} a quota at some point after $\order$ is:
\[
 \mathrm{QuotaCost}_{\order,\texttt{B}} \gets \max \{0, \quota - \tallymax{\order}{\texttt{B} \prec *,r=3}\} = \max \{0, 308 - 120\} = 188
\]
To compute $\mathrm{LeftAtEndCost}_{\order,\texttt{B}}$, a lower bound on the manipulation required to make sure \texttt{B} is not eliminated before $L_{\election,\order} = 3 - (3 - 2) = 2$ other remaining candidates, we need to compute $\mathrm{DispCost}_{\order,\texttt{B},\texttt{A}}$ $=$ $\max \big\{0,\, \frac{1}{2}\big(\tallymin{\order}{\texttt{A},r=3} - \tallymax{\order}{\texttt{B} \prec \texttt{A},r=3} \big) \big\} = 65$ and $\mathrm{DispCost}_{\order,\texttt{B},\texttt{D}}$ $=$ $\max \big\{0,\, \frac{1}{2}\big(\tallymin{\order}{\texttt{D},r=3} - \tallymax{\order}{\texttt{B} \prec \texttt{D},r=3} \big) \big\}$ $=$ $ \max \big\{0,\, \frac{1}{2}\big(163.627 - 120 \big) \big\} = 43.627$.
\ignore{
\begin{eqnarray*}
\mathrm{DispCost}_{\order,\texttt{B},\texttt{A}} & =&  \max \big\{0,\, \frac{1}{2}\big(\tallymin{\order}{\texttt{A},r=3} - \tallymax{\order}{\texttt{B} \prec \texttt{A},r=3} \big) \big\} = 65 \\
\mathrm{DispCost}_{\order,\texttt{B},\texttt{D}} &=&  \max \big\{0,\, \frac{1}{2}\big(\tallymin{\order}{\texttt{D},r=3} - \tallymax{\order}{\texttt{B} \prec \texttt{D},r=3} \big) \big\} \\
& = &  \max \big\{0,\, \frac{1}{2}\big(163.627 - 120 \big) \big\} = 43.627
\end{eqnarray*}}
If we sort these displacements costs in ascending order and take the 2nd element ($L_{\election,\order} = 2$), we find that $\mathrm{LeftAtEndCost}_{\order,\texttt{B}} = 65$ and $\lbdisp_{\order,\texttt{B}} = \max \{ 65, $ $ \min \{ 188, 65 \}\} = 65$. Repeating this process to compute $\lbdisp_{\order,\texttt{D}}$, we find that: $\mathrm{DispCost}_{\order,\texttt{D}} = 24.02$, $\mathrm{QuotaCost}_{\order,\texttt{D}} = 106.04$,  and $\mathrm{LeftAtEndCost}_{\order,\texttt{D}} = 24.02$.
Consequently, $\lbdisp_{\order,\texttt{D}} = 24.02$. The overall displacement lower bound for $\order$ is the minimum of $\lbdisp_{\order,\texttt{B}}$ and $\lbdisp_{\order,\texttt{D}}$, by \autoref{eq:LDISP}. Thus, $\lbdisp_{\order} = 24.02$.

\ignore{
DispCost_{\order,\texttt{D}} \gets \min_{w \in \{A\}} \max \{0, \frac{1}{2}(\tallymin{\order}{w,r=3} - \tallymax{\order}{\texttt{D} \prec w,r=3})\}

\tallymin{\order}{\texttt{A},r=3} = 250
\tallymax{\order}{\texttt{D} \prec \texttt{A},r=3} =  201.96

DispCost_{\order,\texttt{D}} = 24.02

\mathit{QuotaCost}_{\order,\texttt{D}} \gets \max \{0, \quota - \tallymax{\order}{\texttt{D} \prec *,r=3}\} = \max \{0, 308 - 201.96\} = 106.04

\mathit{DispCost}_{\order,\texttt{D},\texttt{A}} = \max \big\{0,\, \frac{1}{2}\big(\tallymin{\order}{\texttt{A},r=3} - \tallymax{\order}{\texttt{D} \prec \texttt{A},r=3} \big) \big\} = \max \big\{0,\, \frac{1}{2}\big(250 - 201.96 \big) \big\} = 24.02

\mathit{DispCost}_{\order,\texttt{D},\texttt{B}} = \max \big\{0,\, \frac{1}{2}\big(\tallymin{\order}{\texttt{B},r=3} - \tallymax{\order}{\texttt{D} \prec \texttt{B},r=3} \big) \big\} = \max \big\{0,\, \frac{1}{2}\big(120 - 201.96 \big) \big\} = 0

\mathit{LeftAtEndCost}_{\order,\texttt{D}} = 0

\lbdisp_{\order,\texttt{D}} = \max \{ 24.02, \min \{ 106.04, 0 \}\} = 24.02

}

\end{example}

\ignore{
\begin{table}[t]
\caption{Comparison of the revised vs original lower bounding heuristics on three example prefixes for the STV election of \autoref{tab:EGSTV1}.}
\centering
\setlength{\tabcolsep}{10pt}
\begin{tabular}{l|c|c|c}
\toprule
 & \multicolumn{3}{c}{\textsc{Prefix $\order$}} \\ 
 \cline{2-4}
 \textsc{Heuristic}   & $[(\texttt{C},1),(\texttt{B},1)]$ & $[(\texttt{C},1),(\texttt{D},0)]$ & $[(\texttt{C},1),(\texttt{A},1)]$ \\
\toprule
\OLD{} Elimination LB &  & & \\
\NEW{} Elimination LB & & & \\
\midrule
\OLD{} Quota LB &  & & \\
\NEW{} Quota LB &  & & \\
\midrule
\NEW{} Displacement LB & & & \\
\bottomrule
\end{tabular}
\end{table}
}

\subsection{New Pruning Methods}\label{sec:NewPruningMethods}

We introduce a \emph{dominance rule} to  maximise the portion of the alternate-outcome search space \NEW{} can ignore. 
We say that node $(l, \order)$ is dominated by node $(l', \order')$ if $l' \leq l$ and the prefixes have the same super-candidate relaxed representations, $\tilde{\order} \equiv \tilde{\order}'$. When deciding whether to add an $(l, \order)$ to our frontier, $F$,  we check whether $(l, \order)$ is dominated by another node already in $F$, or one that we have expanded before. If so, we do not add it to the frontier.

If we have seen an order, $\order'$, with a given relaxed structure, $\tilde{\order}'$, in the past, and we see that structure again in order $\order$, \emph{and} we know the lower bound we attached to  $\order'$, $l'$, is smaller or equal to the lower bound we have attached to $\order$, $l$, then we know that the smallest lower bound we could find for any descendent of $\order'$ will be less than or equal to the smallest lower bound we could find for any descendent of $\order$. The MINLP that 
we create when we add a given sequence of events $\order^*$ to the end of either $\order$ or $\order'$ will be the same. Moreover, the contribution of each elimination or election event to the evaluation of the bound is not dependent on the precise order of prior elimination subsequences.

\section{Results}

\begin{table}[p]
\setlength{\tabcolsep}{2pt}
\caption{For STV elections with $N$ winners,
    $|\mathcal{C}|$ candidates, $|\mathcal{B}|$ validly cast votes, and quota
    $\mathcal{Q}$, we report: margin upper bound (\textbf{UB}), lower
    bounds by various methods, and their run times.
    \OLD{} refers to results published by
    \citep{blom2019toward}, and \OLDR{} to results obtained by  \NEW{}
    without the algorithmic improvements (i.e., a modern implementation of
    \OLD{}). The best margin and time between \NEW{} and \OLDR{} are in bold.
    A `**' indicates that the 10ks limit was reached;
    and `--' that no result was computed for that method.
    Elections where we computed the exact margin only with \NEW{} are shown in bold.}
\centering
\scalebox{0.90}{ 
\begin{tabular}{lrrrrrrrrrr}
\toprule
& & & & & & \multicolumn{3}{c}{\textbf{Margin LB}} &
            \multicolumn{2}{c}{\textbf{Time (sec)}}  \\
\cmidrule(r){7-9}
\cmidrule(l){10-11}
\textbf{Election} & $N$ & $|\mathcal{C}|$ & $|\mathcal{B}|$ & $\mathcal{Q}$ &
\textbf{UB} & \OLD{} & \OLDR{} & \NEW{} & \OLDR{} & \NEW{} \\
\midrule
\multicolumn{11}{l}{\textit{Australian Senate}} \\
ACT'13 & 2 & 27 & 246742 & 82248 & 13391 &   32 &     80 & \bf  2155 & ** & **\\
ACT'16 & 2 & 22 & 254767 & 84923 & 18835 &  224 &    343 & \bf  9162 & ** & **\\
ACT'19 & 2 & 17 &  95749 & 90078 & 12939 &   -- &   1635 & \bf  9933 & ** & **\\
ACT'22 & 2 & 23 & 285217 & 95073 & 11078 &   -- &     85 & \bf   319 & ** & **\\
ACT'25 & 2 & 14 & 293474 & 97825 & 23121 &   -- &  12358 & \bf 16722 & ** & **\\
\bf
NT'13  & 2 & 24 & 102027 & 34494 &  2298 &   96 &    212 & \bf  2298 & ** & \bf 1.12\\
NT'16  & 2 & 19 & 102027 & 34010 & 11244 & 3105 &   3120 & \bf  7438 & ** & **\\
NT'19  & 2 & 18 &  37869 & 35010 & 15890 &   -- &   3563 & \bf  9322 & ** & **\\
NT'22  & 2 & 17 & 103617 & 34540 & 11412 &   -- &    474 & \bf  5092 & ** & **\\
NT'25  & 2 & 17 & 106807 & 35603 & 13256 &   -- &    958 & \bf  5870 & ** & **\\
\addlinespace
\multicolumn{11}{l}{\textit{Minneapolis, USA}} \\
MN BET'09 & 2 & 7 & 16727 & 10696 &  2098 & -- &  2097 & \bf 2098 & 1.38 & \bf 0.67\\
MN BET'13 & 2 & 5 & 23949 & 16286 &  6713 & -- &  6712 & \bf 6713 & 0.62 & \bf 0.21\\
MN BET'17 & 2 & 4 & 48163 & 23232 & 16863 & -- & 16863 &    16863 & 0.31 & \bf 0.10\\
MN BET'21 & 2 & 5 & 42672 & 31876 &  2703 & -- &  2703 &     2703 & 0.4  & \bf 0.17\\
\addlinespace
\multicolumn{11}{l}{\textit{2002 Irish General Election}}\\
Dublin N. & 4 & 13 & 43942 &  8789 &  211 & 189 & 211 & 211 & \bf 165 & 174\\
Dublin W. & 3 &  9 & 29988 &  7498 &  366 & 260 & 366 & 366 & \bf   6 &   8\\
\addlinespace
\multicolumn{11}{l}{\textit{2007 Glasgow City Council}}\\
Linn          & 4 & 11 &  9567 & 1914 & 218 & 114 & 218 & 218 &    287 & \bf  82\\
Newlands      & 3 &  9 &  8654 & 2164 &  88 &  78 &  85 &  85 &      2 &       2\\
Greater P.    & 4 &  9 &  8682 & 1737 & 237 & 147 & 235 & 235 &     67 & \bf  11\\
Craigton      & 4 & 10 & 11052 & 2211 &  75 &  38 &  72 &  72 &      6 &       6\\
Govan         & 4 & 11 &  9560 & 1913 & 309 & 145 & 309 & 309 &   2609 & \bf 468\\
Pollock--s    & 3 &  9 &  9567 & 2392 &   3 &   3 &   3 &   3 & \bf  1 &       2\\
Langside      & 3 &  8 &  2334 & 9334 & 233 & 223 & 233 & 233 &     10 & \bf   5\\
Southside C.  & 4 &  9 &  8738 & 1748 & 229 & 187 & 224 & 224 &    625 & \bf 172\\
Calton        & 3 & 10 &  5199 & 1300 & 376 & 118 & 364 & 364 &    881 & \bf  54\\
Anderston     & 4 &  9 &  6900 & 1381 &  99 &  93 &  99 &  99 &      8 & \bf   7\\
Hillhead      & 4 & 10 &  8984 & 1797 & 105 &  59 & 105 & 105 & \bf 25 &      40\\
Partick W.    & 4 &  9 & 12744 & 2549 & 193 & 185 & 193 & 193 &      5 & \bf   4\\
Garscadden    & 4 & 10 & 10160 & 2033 & 396 & 172 & 396 & 396 &   2310 & \bf 642\\
Drumchapel    & 4 & 10 &  8680 & 1737 & 443 & 253 & 443 & 443 &   3222 & \bf 319\\
Maryhill      & 4 &  8 &  9901 & 1981 & 321 & 223 & 321 & 321 &     62 & \bf  26\\
Canal         & 4 & 11 &  8624 & 1725 & 126 &  98 & 125 & 125 &     11 & \bf  10\\
\bf
Springburn    & 3 & 10 &  5410 & 1353 & 528 &  74 & 505 & \bf 528 &   ** & \bf  441\\
East Centre   & 4 & 13 &  9078 & 1816 & 139 &  65 & 134 &     134 & 2027 & \bf  488\\
\bf
Shettleston   & 4 & 11 &  8803 & 1761 & 353 & 136 & 270 & \bf 353 &   ** & \bf 6170\\
Baillieston   & 4 & 11 & 10376 & 2076 & 105 &  65 & 104 &     104 &    8 & \bf    7\\
North East    & 4 & 10 &  8363 & 1673 & 421 & 251 & 420 &     420 &  813 & \bf  119\\
\bottomrule
\end{tabular}}
\label{tab:ImprovementsPart2}
\end{table}

We report the performance and margin lower bounds obtained by \NEW{} across ten 2-seat Australian Senate elections, four US-based 2-seat STV elections, two Irish STV elections of 3 and 4 seats, and 21 3- or 4-seat Glasgow city county elections from 2007.
We treated each Australian Senate election as a WIGM contest for the purposes of this evaluation.
Our results are in \autoref{tab:ImprovementsPart2}.

The 2013--16 Australian Senate, Irish and Scottish contests were used to evaluate \OLD{} in \citet{blom2019toward}. We use \OLD{} to denote the results published in that paper, and \OLDR{} for results obtained using our re-implementation of their algorithm. We must emphasise that the original implementation (\OLD{}) was designed for a slightly different variant of STV (Australian Senate rules) and not for WIGM. The difference lies in how the transfer value for ballots distributed as part of a winner's surplus is computed. There is no difference in transfer value computation for the first elected candidate across rule sets. The differences arises in any subsequent surplus transfer.
The two variants do not result in different outcomes on any of the contests in \autoref{tab:ImprovementsPart2}, with candidates elected and eliminated in the same order, or substantially different tallies per  round ($\leq$ 10 votes). We expect that the margins for each contest under Senate and WIGM rules would likely be similar. In fact, WIGM and Senate rules are equivalent for 2-seat STV elections as only one surplus distribution is made. This means that tabulation of the Australian Senate and Minneapolis (MN) contests under WIGM and Senate rules is equivalent. The original results are included to give a sense of progress due to advances in solving technology, modern computational power, and improvements to the algorithm itself.

All experiments were run on a server with 32 virtual CPUs and 128~GB of RAM.
A 10,000s time limit was applied for each run of \OLDR{} and \NEW{}. \citet{blom2019toward} applied a 12-hour time limit when generating their results (\OLD{}). All runs of \OLDR{} and \NEW{} use  parallelisation when expanding nodes on the frontier, expanding up to 30~nodes simultaneously.

\section{Concluding Remarks}

The improvements we have made to the STV margin lower bounding method of \citet{blom2019toward} have resulted in substantially tighter lower bounds for large real-world STV elections than previously computable. For smaller contests, including many of the non-Australian cases in our dataset, we are able to compute exact margins or near-exact lower bounds in substantially less time. The lower bounds we compute in these cases are sufficient for use in mismatch-based RLAs \cite{ek2025doing}. 
For example, for the NT 2022 election assuming a mismatch rate of 0.1\%, a mismatch-based audit would require sampling on average about 55 ballots based on the STV-26 lower bound, but more than 1000 ballots based on the BST-19* one.
If the mismatch rate was 1\% then these numbers jump to
106 and almost 100,000 respectively.
Further work is required to scale the method for use on large Australian Senate contests with hundreds of candidates and 6 to 12 seats.


\bibliographystyle{splncsnat}
\bibliography{references}


\newpage
\appendix

\section{ConcreteSTV Upper Bounds Details}\label{sec:concrete}

We ran ConcreteSTV with settings that matched, as closely as possible, the variant chosen in this paper. However, it is always possible 
that slight differences mean that the bounds found with one version are not guaranteed to be valid in a different version.\footnote{For example,
the lower-bounding algorithm uses floating-point values; the ConcreteSTV implementation uses fixed-precision decimals with 6 decimal places. This is
unlikely to make any practical difference, but in principle it could.} 
The margin lower bounding algorithm that we present in this paper is not reliant on the correctness of provided upper bounds. If the provided upper bound is less than the true margin, the algorithm will return a lower bound that is less than or equal to it. This is still a valid lower bound on the margin, although the algorithm may have been able to find a better lower bound with a valid initial upper bound.

\section{Minimal manipulations MINLP}\label{app:minlp}

We present a MINLP designed to find a minimal manipulation of an STV election such that a specific partial or complete election order $\order$ is realised. This model assumes the use of Weighted Inclusive Gregory STV. We define a ballot type as a specific ranking over candidates that may appear on a cast ballot. 

\subsection{Indices, Sets, Parameters}
\begin{longtable}{p{25pt}p{300pt}}
$\mathcal{B}$ & Ballots cast in the original election profile. \\[2pt]
$c, \mathcal{C}$ & Candidates. \\[2pt]
$s, \mathbb{S}$ & Ballot types (or signatures). \\[2pt]
$N_s$ & Number of ballots of type $s \in \mathbb{S}$ cast in the original election profile. \\[2pt]
$r, \mathcal{R}$ & Rounds of tabulation. \\[2pt]
$L$ & Last round in which a candidate is either eliminated or elected to a seat with a quota in $\pi$.\\[2pt]
$Q$ & Quota. \\[2pt]
$A_r$ & The subset of candidates still standing at round $r$ of $\pi$ \\[2pt]
$S$ & Number of available seats.\\[2pt]
\end{longtable}

\subsection{Variables}
All non-binary variables are continuous in this model. This is a slight relaxation.

\begin{longtable}{p{25pt}p{300pt}}
$p_s$      & Number of ballots that are modified so that their new type is $s \in \mathbb{S}$. \\[2pt]
$m_s$      & Number of ballots whose original type is $s \in \mathbb{S}$ but have now been changed to a different type. \\[2pt]
$y_s$      & Number of ballots of type $s \in \mathbb{S}$ cast in the new election profile. \\[2pt]
$v_{c,r}$  & Tally of candidate $c$ at the start of round $r$. \\[2pt]
$q_{c,r}$  & Binary variable with value 1 iff the tally of candidate $c$ at the start of round $r$ is at least a quota, and 0 otherwise. \\[2pt]
$nq_{c,r}$ & For convenience, we define a binary $nq_{c,r}$ whose value is 1 iff the tally of candidate $c$ at the start of round $r$ is less than a quota. \\[2pt]
$t_{r}$    & Transfer value applied to ballots leaving an elected candidates' tally in round $r$. These variables are only defined for rounds where a candidate has been seated after achieving a quota, and their ballots distributed at a reduced value. \\
\end{longtable}

\subsection{Functions}

For each candidate $c$, and round $r$ of $\pi$, we define $f(\pi, c, r)$ as returning a list of tuples ($s$, $v$, $Caveats$) where $s$ denotes a ballot type, $v$ denotes the value of each ballot of that type to $c$, assuming the conditions in $Caveats$ hold, and $Caveats$ a list of binary $q_{c',r'}$ and $nq_{c',r'}$ variables whose values must equal 1 for $c$ to be awarded ballots of type $s$, each with value $v$, in round $r$. If a ballot moves from eliminated candidate to eliminated candidate before it reaches $c$ in $r$, it's value will be 1 ($v = 1$) and $Caveats$ empty. For example, consider the ranking $s$ $=$ ($\texttt{A}$, $\texttt{B}$, $\texttt{C}$) and the order $\pi$ $=$ [($\texttt{A}$, 0), ($\texttt{D}$, 0), ($\texttt{B}$, 0)]. The function $f(\pi, \texttt{C}, 2)$ will return a set of tuples that includes ($s$, 1, []). 

If we know that a ballot will have formed part of one or more surplus transfers before it reaches $c$ in $r$, then its value will equal the product of these transfer values. For example, consider the ranking $s$ $=$ ($\texttt{A}$, $\texttt{B}$, $\texttt{C}$) and the order $\pi$ $=$ [($\texttt{A}$, 1), ($\texttt{D}$, 0), ($\texttt{B}$, 0)], in which $\texttt{A}$'s transfer value was 0.125. The function $f(\pi, C, 2)$ will return a set of tuples that includes ($s$, 0.125, []). For the ranking $s$ $=$ ($\texttt{A}$, $\texttt{F}$, $\texttt{C}$) and order $\pi$ $=$ [($\texttt{A}$, 1), ($\texttt{D}$, 0), ($\texttt{F}$, 1), ($\texttt{B}$, 0)], with $\texttt{A}$ and $\texttt{F}$'s transfer values being 0.125 and 0.05, respectively, the function $f(\pi, \texttt{C}, 3)$ will return a set of tuples that includes ($s$, 0.00625, []).

$Caveats$ will be non-empty in situations where the ballot could have skipped over an elected candidate $c'$ on it's way to $c$, due to $c'$ already having a quota. For  $s$ $=$ ($\texttt{A}$, $\texttt{F}$, $\texttt{C}$) and order $\pi$ $=$ [($\texttt{A}$, 1), ($\texttt{F}$, 1), ($\texttt{B}$, 0)], with $A$ and $F$'s transfer values being 0.125 and 0.05, respectively, the function $f(\pi, \texttt{C}, 2)$ will return a set of tuples that includes both ($s$, 0.00625, [$nq_{F,1}$]) and ($s$, 0.125, [$q_{F,1}$]).

\subsection{Objective}
We minimise the number of ballots modified:
\begin{equation}
    \min \quad \sum_{s} p_s
\end{equation}

\subsection{Constraints}

The number of ballots cast of type $s \in \mathbb{S}$ in the manipulated election profile is equal to the number of ballots originally cast of that type ($N_s$) in addition to the number of ballots of other types modified to have type $s$ ($p_s$), minus the ballots of type $s$ in the original profile changed to a different type ($m_s$). 
\begin{align}
y_s & =  N_s + p_s - m_s & \\
\sum_s p_s & =   \sum_s m_s & 
\end{align}
For candidates $c$ that are elected to a seat in $\pi$ at a round $r' \leq L$:
\begin{align}
v_{c,r} & \geq  Q  q_{c,r} & \forall r < r'  \\
v_{c,r} & \leq  (1 - q_{c,r}) (Q - \epsilon) + |\mathcal{B}| q_{c,r} &  \\
q_{c,r'} & =   1 & 
\end{align}
For rounds $r < L$ in which a candidate $c$ is elected to a seat in $\pi$:
\begin{align}
    t_{r} v_{c,r} & =  v_{c,r} - Q  & \label{cons:tv}
\end{align}
For candidates $c$ that are eliminated in $\pi$ at a round $r \leq L$:
\begin{align}
v_{c,r} & \leq  Q - \epsilon & \\
v_{c,r} & \leq  v_{c',r} & \forall c' \in A_r \setminus \{c\} 
\end{align}
The following constraints define the number of votes in the tally piles of each candidate $c \in \mathcal{C}$ at the start of each round $r$ ($v_{c,r}$) for all rounds $r$ where $c \in \mathcal{D}_r$. 
\begin{align}
    v_{c,0} & =  \sum_s y_s & \forall c \in \mathcal{C} \label{cons:numballots_0}\\
    v_{c,r} & =  v_{c,r-1} + \sum_{(s,v,C) \in f(\pi,c,r-1)} v \,y_s \prod_{x \in C} x & \forall r \in [1, L], c \in A_r \label{cons:value_ballots}
\end{align}


\end{document}